# Mathematical Modeling of the Spread of COVID-19 in Moscow and Russian Regions

E.M. Koltsova[1*], E.S. Kurkina[1,2], A.M. Vasetsky[1]

[1] Department of Digital Technologies and Chemical Engineering, D. Mendeleev University of Chemical Technology of Russia. Miusskaya sq. 9, Moscow, 125047, Russia.
[2] Department of Computational Mathematics & Cybernetics, Lomonosov Moscow State University, GSP-1, Leninskie Gory, Moscow, 119991, Russia

**Abstract:**

To model the spread of COVID-19 coronavirus in Russian regions and in Moscow, a discrete logistic equation describing the increase in the number of cases is used. To check the adequacy of the mathematical model, the simulation results were compared with the spread of coronavirus in China, in a number of European and Asian countries, and the United States. The parameters of the logistics equation for Russia, Moscow and other large regions were determined in the interval [01.03 – 08.04]. A comparative analysis of growth rates of COVID-19 infected population for different countries and regions is presented. Various scenarios of the spread of COVID-19 coronavirus in Moscow and in the regions of Russia are considered. For each scenario, curves for the daily new cases and graphs for the increase in the total number of cases were obtained, and the dynamics of infection spread by day was studied. Peak times, epidemic periods, the number of infected people at the peak and their growth were determined.

**Keywords:** coronavirus COVID-19, mathematical modeling, logistic equation, epidemic development scenarios.

## 1. Introduction

The discrete logistic equation is use to describe the spread of the epidemic coronavirus COVID – 19 in Russia and its regions. For the first time, the logistic equation in differential form was suggested by the Belgian mathematician Pierre Verhulst in 1845 [1] to model population growth. The essential difference from the mathematical model of Thomas Malthus (presented in the famous work "Experience of the population law" [2]), which describes exponential population growth, is that the Verhulst model took into account competition for resources,

[*]Corresponding author. Email: kolts@muctr.ru

which leads to limited population growth. In 1920 logistic equation in differential form

$$\frac{dy}{dt} = \lambda y(1 - \frac{y}{N}), \qquad (1)$$

where y(t) is the population size at time t, the parameter λ characterizes the population growth rate, and the parameter N determines the maximum possible population size in conditions of limited resources.

The logistic equation began to be widely used, starting from the 20s of the last century, when it was rediscovered by R. Pearl [3] and confirmed its adequacy in experiments with the reproduction of Drosophila flies.

Currently, the use of this equation has found wide application in mathematical Biophysics, which is well reflected in the monographs of Russian biophysicists G. Yu. Riznichenko and A. B. Rubin [4].

The Verhulst equation was also used to describe the spread of epidemics. In this case, the entire population that can become infected is divided into two parts: susceptible to the disease (Susceptible) and infected (-Infectious). It is believed that the disease is transmitted through contacts of healthy people with patients with a probability of λ, and under conditions of good mixing, the increase in patients is described by the equation:

$$\frac{dx}{dt} = \lambda x(1 - x). \qquad (2)$$

Here $x = y/N$ is the part of infected people,

$(1 - y/N)$ is the part of people susceptible to the disease.

N is the maximum number of cases.

This model, written in the form of two equations, is called the *SI* model (*Susceptible-Infectious model*).

Equation (2) is still being used for modeling COVID-19 coronavirus in China [5] and Sweden [6]. It has two fixed (stationary) points: .: $\overline{y_1} = 0$, $\overline{y_2} = N$. The second point is the only attractor. Thus, over time, no matter what value the growth indicator λ in equation (2) takes, the population size will tend to N only with different periods of time to reach this stationary state.

The *SIR* model (*Susceptible-Infectious-Recovered model*) is considered more realistic. In it, the entire population is divided into three parts: the part susceptible to the disease *y* (*Susceptible*), infected *x* (*Infected*) and recovered *z* (*Recovered*). The presence of recovering people in the model reduces the number of infected people who can infect others. The model equations have the form:

$$\frac{dy}{dt} = -\lambda y \frac{x}{N},$$

$$\frac{dx}{dt} = +\lambda y \frac{x}{N} - \mu x,$$

$$\frac{dz}{dt} = +\mu x,$$

where $\mu$ – is the rate of recovery.

There are even more complex models [7] that describe the spread of infections more accurately. However, the more complex the model, the more unknown parameters it contains, the value of which cannot be estimated with good accuracy.

Note, that there is a large difference between the maximum number of residents, who are potentially ill ($N_{max}$) and the maximum number of residents, who are actually ill ($N_{in}$). The value of $N_{max}$ is much greater than the number of $N_{in}$.

The main parameters that most strongly affect on the spread of COVID-19 coronavirus and on the characteristics of the peak are two parameters: the growth rate of the number of cases and the maximum value of the number of residents who can potentially be infected $N_{in}$. The number of recovered in the case of COVID-19 almost up to the peak can be ignored.

The main parameters that most strongly affect the rate of spread of COVID-19 coronavirus infection and the characteristics of the peak are two parameters: the growth rate of the number of cases and the maximum value of the number of residents who can potentially be infected with $N_{in}$.

We used a discrete logistic equation that contains exactly these two parameters, and we believe that in this case it describes the spread of the epidemic better than any other model. The discrete logistic equation is widely known thanks to the work of the American scientist M. Feigenbaum [8]. M. Feigenbaum discovered interesting regularities obtained using this equation and created a theory of universality for discrete maps. In particular, he showed that when the dimensionless growth indicator changes in the equation in the interval (3-4), period doubling bifurcations occur, and when the growth indicator value is equal to ~ 3.5699, the logistic equation generates chaotic dynamics. After these studies, the discrete logistic equation was widely used for modeling various processes [9-10]. For an adequate discrete description of phenomena, an important characteristic is the choice of a step, a discrete time interval, at which the population numbers are considered.

According to calculations by Nobel laureate Michael Levitt, made for China, in 24 hours, 1 person in China could infect 2.2 million people with a coronavirus infection. That is, one person could become infected from another within a time interval of ~ 10.9 hours. Therefore, a time interval of 12 hours was taken for the discrete logistic equation. This period is close to 10.9 hours and is convenient for comparing the estimated and actual data. In other words, the number of patients is recalculated every 12 hours, and the actual data is checked every day.

## 2. Mathematical model of the spread of COVID-19

We used a discrete logistic equation to describe the spread of COVID-19 coronavirus in different countries and cities. It has the form:

$$y_{n+1} = \lambda y_n (1 - y_n/N), \qquad (3)$$

where $y_n = y(t_n), n = 1,2,\ldots,n_{max}$ — is the number of cases in the $n^{th}$ time $t_n$; $\lambda$ is the coefficient (index) of growth of population, which can be varied by changing the conditions of infection, for example, when carrying out quarantine measures; $N$ is the normalization factor (the maximum value of the number of residents who can potentially get infected). This parameter depends on a number of factors, such as the population size, its crowding or density, resistance to disease, discipline of the population during quarantine measures, etc.

This model describes growth in accordance with a logistic function: the number of sick people grows rapidly until it is small ($y_n \ll N$), then the growth rate slows down more and more, and the number of sick people asymptotically tends to a stationary value. Making the change of variables:

This model shows that the population of sick people grows rapidly (exponentially), while it is small ($y_n \ll N$) and begins to decrease, when there are a lot of sick people. Making the change of variables

$$y_n = x_n N, \quad \alpha = \lambda. \qquad (4)$$

we reduce equation (3) to the form:

$$x_{n+1} = \alpha\, x_n (1 - x_n), \qquad (5)$$

where the variables $x_n$ and the parameter $\alpha$ are dimensionless.

For values $0 < \alpha \leq 1$, regardless of the choice $x_0$, the population size tends to zero. That is, the number of cases will tend to zero, no matter how many cases there were at the beginning.

At values $1 < \alpha \leq 3$, the dimensionless number of the diseased population tends to a stationary stable state $\bar{x}$, equal to

$$\bar{x} = \frac{\alpha - 1}{\alpha} \qquad (6)$$

Therefore, over time, the number of people who become ill at the end of the epidemic will be equal to

$$\bar{y} = \left(\frac{\alpha - 1}{\alpha}\right) \cdot N \qquad (7)$$

The relations (6) and (7) are correct if the $\alpha$ index remains constant throughout the epidemic.

Note that there is an important difference between the differential equation describing logistic growth and the discrete one. In the differential equation at $1<\alpha$, the number tends to the value $N$, and in the discrete equation-to the value $\bar{y}$ (7), which depends on the growth rate indicator, and may differ greatly from $N$. The current quarantine measures reduce the probability of infection and the indicator $\alpha$ decreases, which leads to a decrease in the total number of cases $\bar{y}$.

### 3. The results of mathematical modeling

To test the feasibility of using the logistic equation to describe the spread of COVID-19 coronavirus in Russian regions, the model was applied to describe the spread of the virus in China, where epidemic has already ended [11].

The initial value of the normalized multiplier $N$ for China was selected using the ratio (7). For China, the number of cases at the end of the epidemic was $\bar{y} = 82500$. The initial approximation of the average value of the population growth indicator was taken as $<\alpha> \sim 1.12$.

The rates $\alpha$ and the value of the normalizing factor $N$ were determined from the trend of the population curve of COVID-19 infected people throughout the entire period of the epidemic. The time interval $\Delta = t_{n+1} - t_n$, with which the number of infected populations was calculated, was taken to be 12 hours, and comparison with statistical data was carried out after $2\Delta$, that is, once a day. The value of the normalizing factor was chosen $N = 760{,}000$, and the parameters for the growth of the infected population were equal to

$\alpha = 1.19$           from 3.01.2020 to 30.01.2020,
$\alpha = 1.104$         from 31.01.2020 to 10.02.2020,
$\alpha = 1.18$           from 10.02.2020 to 12.02.2020,
$\alpha = 1.119$        from 13.02.2020 till the end of the epidemic.

Figures 1 and 2 show the change in the population of people infected with COVID-19 coronavirus in China and the increase in the number per day for the period of the epidemic.

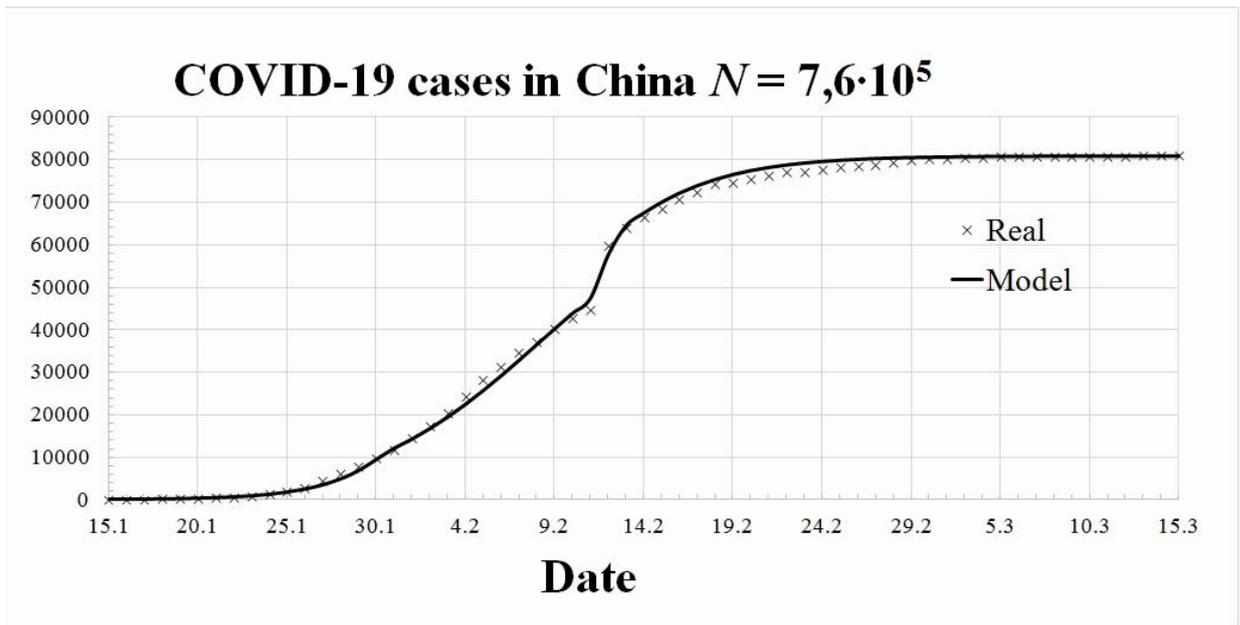

**Fig. 1.** Curves of the total infected cases with COVID-19 coronavirus during the epidemic in China

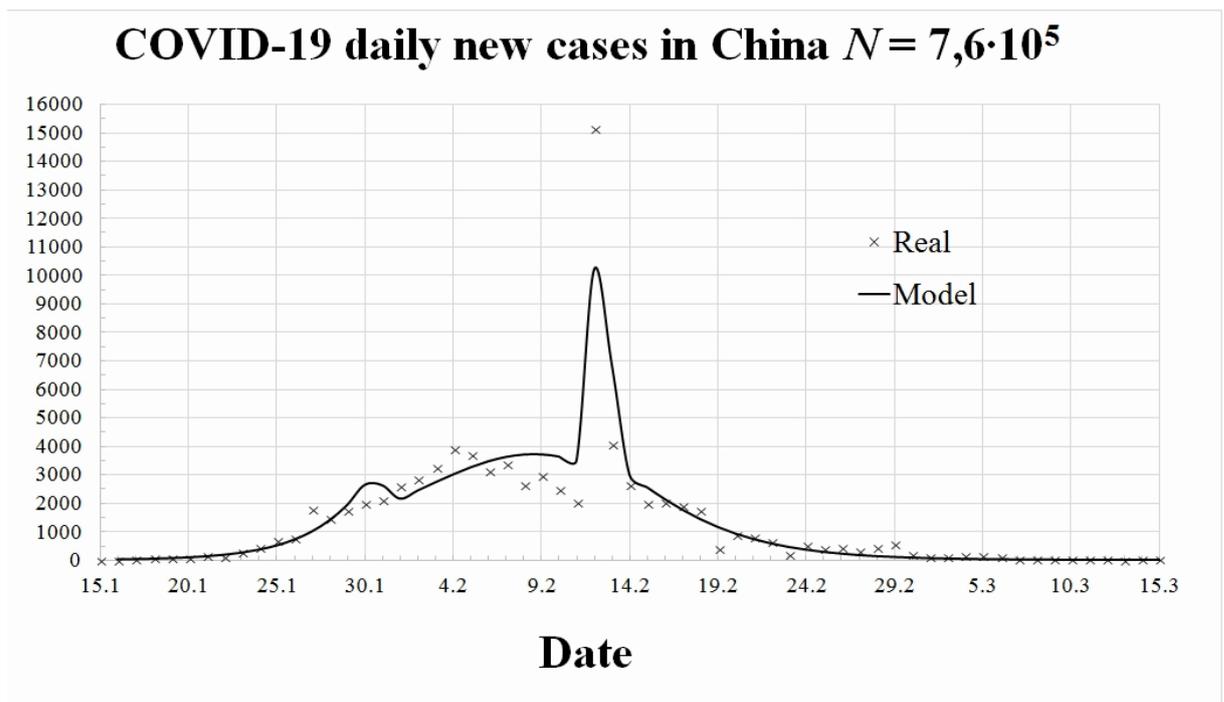

**Fig. 2.** Curves of the daily new cases with COVID-19 coronavirus during the epidemic in China

Figures 1 and 2 shows a good correspondence between the calculated and real data. This good coincidence made it possible to use the logistic equation (3), (5) to model the spread of the coronavirus epidemic in a number of European and Asian countries. For each of the countries, indicators of population growth $\alpha$ and normalizing factors $N$ were found.

Table 1 shows the found average growth rates of the diseased population before the peak of the epidemic in a number of European and Asian countries and the United States.

Table 1. Average growth rates of the number of cases before the peak of the epidemic by country

| № | Country | growth rate α |
|---|---|---|
| 1 | Germany | 1.19 |
| 2 | Portugal | 1.18 |
| 3 | Spain | 1.163 |
| 4 | China | 1.16 |
| 5 | Italy | 1.152 |
| 6 | France | 1.149 |
| 7 | USA | 1.148 |
| 8 | South Korea | 1.121 |
| 9 | Sweden | 1.088 |
| 10 | Japan | 1.047 |

The normalizing factor N was also determined for a number of European and Asian countries. It was found from a comparison of estimated and actual data. The actual data was taken from the site [12].

Table 2. The ratio of the maximum possible number of people potentially infected with the virus in a city or country to the population

| № | Country, city | N/A·100% (A – number of cities, countries) |
|---|---|---|
| 1 | New-York, USA | 11,6 |
| 2 | Wuhan, China | 4,75 |
| 3 | Sweden | 5 |
| 4 | Spain | 4,4 |
| 5 | Italy | 3,66 |
| 6 | Portugal | 3 |
| 7 | South Korea | 1,96 |
| 8 | Germany | 1,34 |
| 9 | USA | 1,2 |
| 10 | Japan | 0,55 |
| 11 | China | 0,053 |

After analyzing the ratio of the normalizing factor to the number of residents of Moscow, four scenarios for the development of the spread of the COVID-19 coronavirus epidemic were considered:

1 scenario:   N = 1 000 000  (N/A·100% = 7,1 %);        (8.1)
2 scenario:   N = 760 000    (N/A·100% = 5,4 %);         (8.2)
3 scenario:   N = 500 000    (N/A·100% = 3,6 %);         (8.3)
4 scenario:   N = 300 000    (N/A·100% = 2,1 %).         (8.4)

Figures 3 and 4 show actual data (from 05.03.2020 to 09.04.2020) and calculated distribution curves for the number of cases and their growth over all days of the epidemic (the forecast).

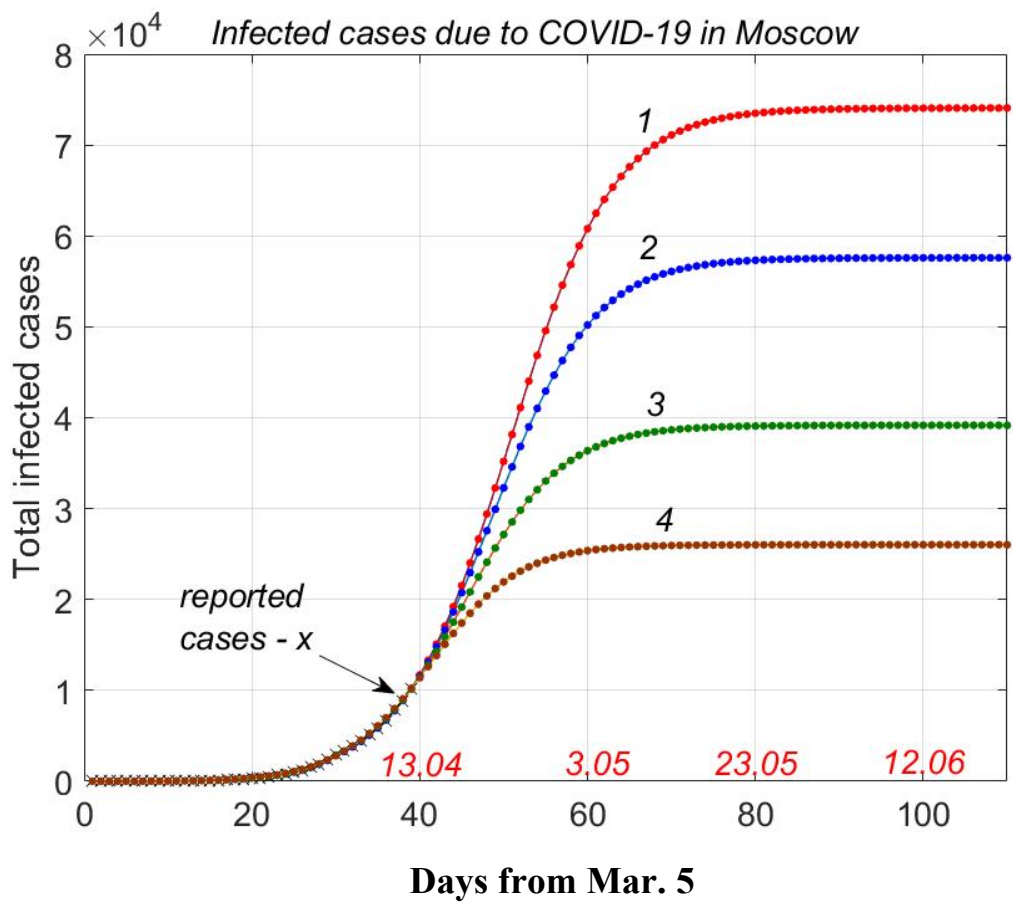

**Days from Mar. 5**

**Fig. 3**. Curves of the number of infected people by epidemic days in accordance with the four scenarios (1 – 4).

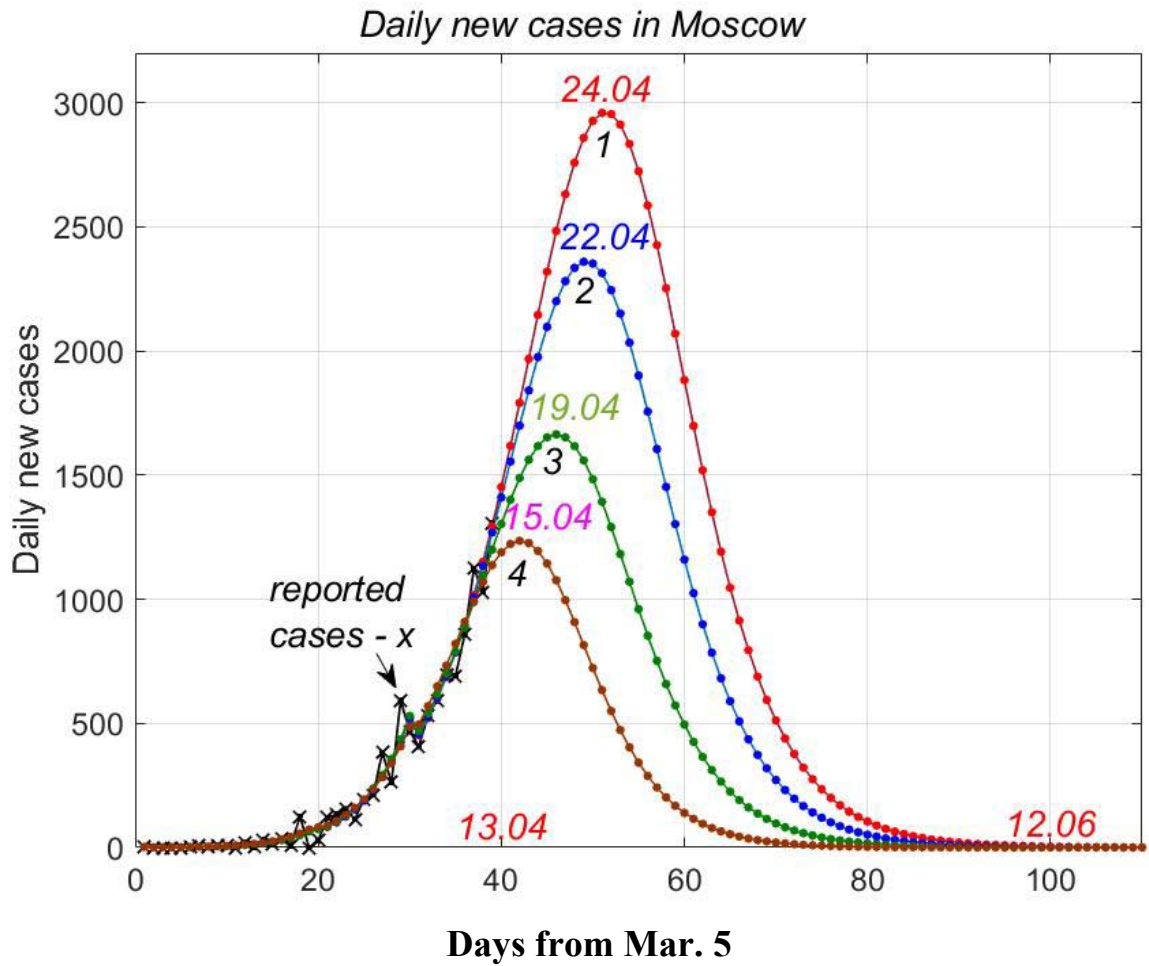

**Fig. 4**. Curves of the daily new cases by epidemic days in accordance with the four scenarios (1 – 4).

The values of growth rates of the diseased population for fourth scenarios are presented in Table 3.

**Table 3. The growth rates of the number of cases for 4 scenarios of epidemic spread in Moscow**

| $N$ | $10^6$ | $7.6\times10^5$ | $5.0\times10^5$ | $3.0\times10^5$ |
|---|---|---|---|---|
| $\alpha$ | $\alpha_1$ | $\alpha_2$ | $\alpha_3$ | $\alpha_4$ |
| from March 5 to March 21 | 1.14 | 1.14 | 1.14 | 1.143 |
| From March 22 to March 31 | 1.111 | 1.111 | 1.113 | 1.108 |
| from April 1 till the end of the epidemic | 1.08 | 1.1082 | 1.085 | 1.095 |

Also, simulations of the spread of the COVID-19 coronavirus epidemic in the regional part of Russia was done. (The number of infected residents of the

Moscow region and Moscow was subtracted from the total number of infected people in Russia). Table 4 shows the growth rates of the number of cases for various scenarios with a normalizing factor:

$$N = 2\cdot 10^6 - 1^{st} \text{ scenario,}$$
$$N = 1\cdot 10^6 - 2^{nd} \text{ scenario,}$$
$$N = 7,6\cdot 10^5 - 3^{th} \text{ scenario,}$$
$$N = 5\cdot 10^5 - 4^{th} \text{ scenario.}$$

**Table 4. The growth rates of the number of cases for 4 scenarios of the spread of coronavirus epidemic in the regional part of Russia**

| Normalizing factor, N | Growth rates, $\alpha$ | Retention time for $\alpha$ |
|---|---|---|
| 1 scenario: $N = 2\cdot 10^6$ | 1.13 | 01.03-14.03 |
|  | 1.098 | 15.03-30.03 |
|  | 1.0865 | 31.03-… |
| 2 scenario: $N = 1\cdot 10^6$ | 1.13 | 01.03-14.03 |
|  | 1.098 | 15.03-30.03 |
|  | 1.0875 | 31.03-… |
| 3 scenario: $N = 7,6\cdot 10^5$ | 1.13 | 01.03-14.03 |
|  | 1.098 | 15.03-30.03 |
|  | 1.088 | 31.03-… |
| 4 scenario: $N = 5\cdot 10^5$ | 1.13 | 01.03-14.03 |
|  | 1.098 | 15.03-30.03 |
|  | 1.0895 | 31.03-… |

Figure 5 shows actual and estimated data on the number of people infected with COVID-19 coronavirus per day for four scenarios.

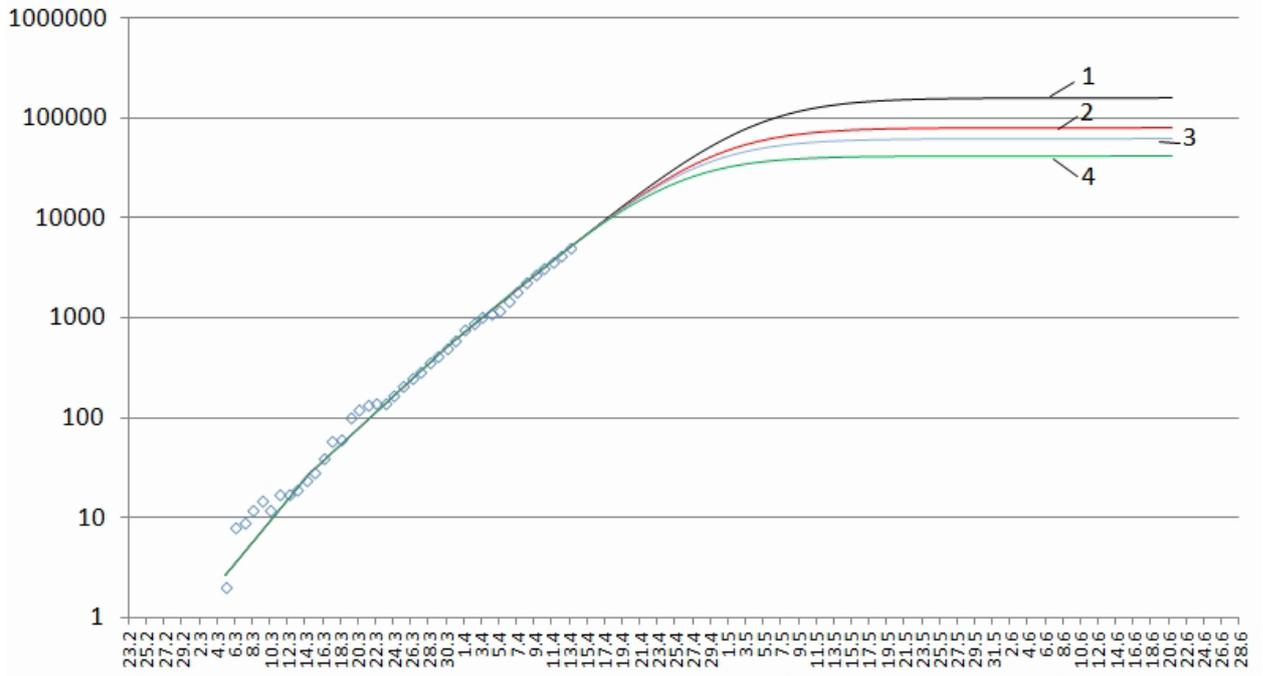

**Fig. 5**. Curves of the number of infected people in the regional part of Russia by epidemic days in accordance with the four scenarios (1 – 4), ◊ - reported cases.

Figure 6 shows the actual and estimated daily growth of people infected with COVID-19 coronavirus for four scenarios during the epidemic.

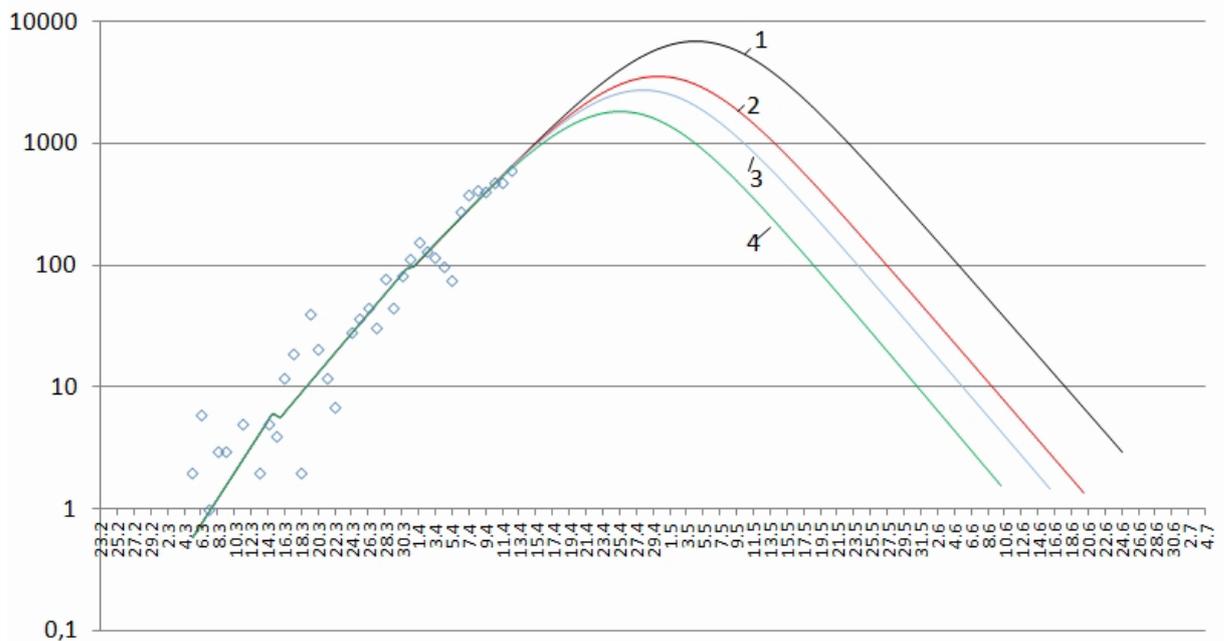

**Fig. 6**. Curves of the daily new cases in the regional part of Russia by epidemic days in accordance with the four scenarios (1 – 4), ◊ - reported cases.

Table 5 shows the calculated data under four scenarios: the time of peak population growth, the number of infected people and the value of new case at the peak, the total number of cases, and the end time.

**Table 5. Calculated data of peculiar parameters of the COVID-19 coronavirus epidemic in regional Russia under 4 scenarios**

| Normalizing factor N | Peak time | Cases at the peak | Daily cases at the peak | Total number of cases | End time of the epidemic |
|---|---|---|---|---|---|
| 1 scenario: $N = 2 \cdot 10^6$ | May 4 | 85579 | 6957 | 160444 | early July |
| 2 scenario: $N = 1 \cdot 10^6$ | April 30 | 44666 | 3542 | 80788 | mid-late June |
| 3 scenario: $N = 7,6 \cdot 10^5$ | April 28 | 35562 | 2729 | 61768 | early-mid June |
| 1 scenario: $N = 5 \cdot 10^5$ | April 25 | 21934 | 1845 | 41165 | early June |

The growth rates of the number of people infected with coronavirus in the Russian regions were calculated in such a way as to best describe the actual data from March 1 to April 7. We see that they changed twice. Since we use only recorded data, the change in the indicator for the first time is probably due to an increase in the number of tests and an improvement in their quality.

The decrease in the indicator since April 1st is due to the reaction of the virus spread to quarantine measures Then, starting from April 8th, a forecast was made using equation (3), which used the value of the last growth indicator.

Table 6 shows a comparison of forecast and real data for several days (from 15.04.20 to 21.04.20) for the same four scenarios.

Table 6. Comparison of real and calculated data

| Day | Reported cases | 1 scenario: $N = 2 \cdot 10^6$ | 2 scenario: $N = 1 \cdot 10^6$ | 3 scenario: $N = 7{,}6 \cdot 10^5$ | 4 scenario: $N = 5 \cdot 10^5$ |
|---|---|---|---|---|---|
| April 15 | 7127 | 7456 | 7425 | 7305 | 7185 |
| April 16 | 8738 | 8774 | 8681 | 8477 | 8277 |
| April 17 | 10358 | 10218 | 10131 | 9807 | 9481 |
| April 18 | 12111 | 11985 | 11739 | 11303 | 10829 |
| April 19 | 13941 | 13915 | 13323 | 12975 | 12287 |
| April 20 | 15647 | 16190 | 15324 | 14825 | 13857 |
| April 21 | 17402 | 18793 | 17554 | 16853 | 14768 |

Analyzing the data in the table 6, you can see that the best agreement with the actual data is given by calculations for the first and second scenarios with the parameter $N = 2 \times 10^6$ and $N = 1 \times 10^6$. But the final choice of the normalizing factor $N$ can be made closer to the peak value of the daily new cases.

## 4. Discussion

The simplest model of infection spread chosen by us, which is based on a discrete logistic equation (3), has demonstrated a very good description of real data and the ability to predict the dynamics of the epidemic. Based on the data on the spread of COVID-19 coronavirus infection in China, where the epidemic has already ended, we worked out the procedure for finding the model parameters. The model was used to describe the dynamics of the spread and forecast of COVID-19 coronavirus infection in various countries and cities: Spain, Italy, France, Portugal, Sweden, South Korea, Japan, the United States, and others. For each country, different model parameters were selected at the early stages of the spread of the epidemic and several scenarios for its further development were proposed. All scenarios described the actual data well at the beginning of the epidemic and produced very similar results. But as the extent of the epidemic increases, when the influence of the normalizing factor $N$ begins to affect, the forecast scenarios for the development of the epidemic begin to differ greatly. Then the final value of the parameter $N$ can be selected based on the actual data.

Note, that the rate of increase in the number of cases is evaluated in the first stage of the epidemic, when the linear term prevails in equation (5), and the nonlinear term can be ignored:

$$\frac{x_{n+1}}{x_n} \approx \alpha \qquad (9)$$

Then the disease increases exponentially, which in the logarithmic scale is a straight line. The growth indicator determines the slope of the straight line. If the growth rate changes, then the slope of the straight line also changes. It may change with the introduction of various quarantine measures or their cancellation.

Table 1 shows the average estimated growth rates of the number of cases before the peak of the epidemic in a number of European and Asian countries and the United States. The best agreement with actual data was obtained for these values of growth rates. These values determined both the time of the peak and the number at the peak. Table 1 shows that Sweden and Japan have the lowest rates of growth in the number of cases. Perhaps because of these indicators, these countries did not impose strict quarantine measures.

As mentioned above, the parameter $N$ begins to affect strongly on the development of the epidemic closer to its peak. Although this parameter represents the maximum number of residents who can potentially be infected, it does not mean that all of them will get sick. In this model, as much as corresponds to the growth indicator $\alpha$ will get sick in contrast to *SI, SIR,* and other models. The normalizing factor $N$ depends on the nation's immunity to the virus, on the living conditions of the inhabitants (crowding, etc.), on the mentality of the nation, etc. It is calculated based on the agreement of calculated and actual data. Table 2 shows the ratio of $N$ to population $A$ of the cities or countries considered (N/A·100%).

Let's consider possible scenarios for the development of the epidemic in Moscow. The beginning of the spread of coronavirus infection in Moscow is considered to be March 5, when 5 cases were identified. To determine the normalizing factor $N$ for Moscow, Table 2 was analyzed and four possible scenarios for the spread of COVID-19 in Moscow (8.1) – (8.4) with a ratio (N/A) of 2 to 7% were proposed.

The first scenario (8.1) is the most hard, let's call it New York scenario, the second scenario (8.2) is called Wuhan (the normalizing factor coincides with the normalizing factor taken for Wuhan). The last fourth scenario (8.4) is the easiest, let's call it the Israeli one (the normalizing multiplier is the same as in Israel).

The parameter $\alpha$ for Moscow was selected at the early stages of infection spread (before April 7). Fig. 7 shows the increase in the number of cases by day in a logarithmic scale (the black line marked with crosses is statistical data, the red curve is model calculations for the scenario with N = 760000). We see that up to the last day of observation (April 18), the model perfectly describes the statistical data. The graph of the number of cases shows that the entire observation segment can be approximately divided into three intervals with different indicators $\alpha$, in which the growth trend of patients has a different slope. The values of the indicator $\alpha$ are shown in Fig. 7 and in the Table 3. We see that the value of the indicator decreased when measures for self-isolation of people gave results. In Moscow, these measures had an impact on April 1 (Fig. 7. the slope of the straight line decreased in a logarithmic scale).

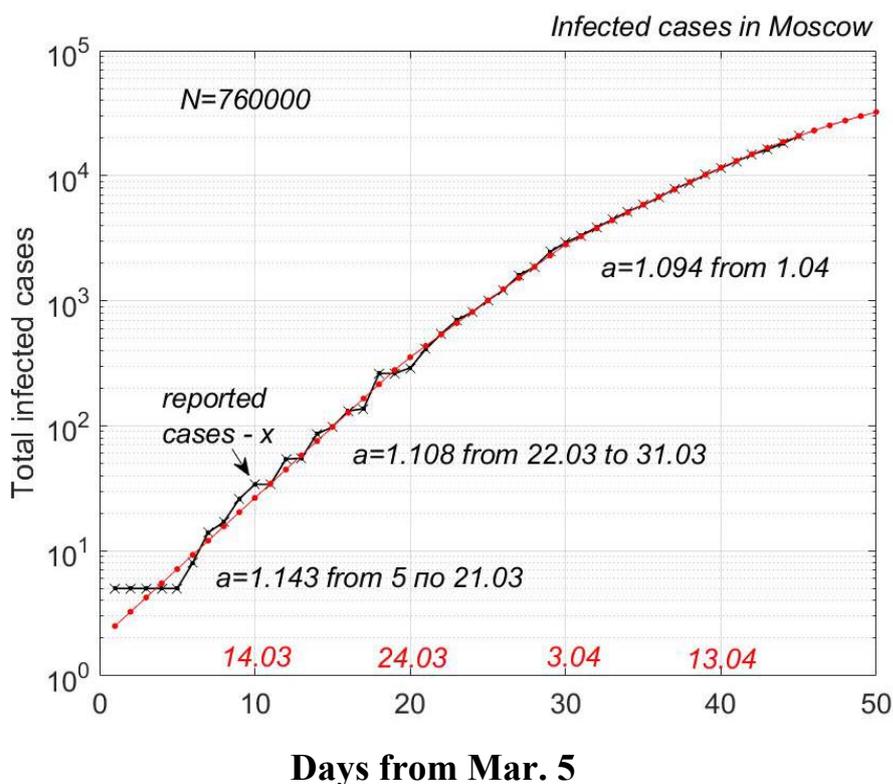

**Days from Mar. 5**
**Fig. 7.** Determination of the growth index $\alpha$ of the number of patients with COVID-19 coronavirus in Moscow

The parameter values have not been re-selected or changed since April 7. All scenarios up to April 7 described the actual data well, but by April 14 it became clear that the 3$^{rd}$ and 4$^{th}$ scenarios were lagging behind the actual data, especially the 4-th scenario with N = 3×10$^5$. Splitting of forecast data under four scenarios begins as the number of cases increases and approaches the peak (see Fig. 3, 4).

Table 5 shows the calculated data under four scenarios for the spread of the coronavirus epidemic in Moscow: the time of peak population growth, the number of infected people and the value of new case at the peak, the total number of cases, and the end time. The last column shows the estimated time of the end of the epidemic (when only a few people get sick a day).

Table 8 shows a comparison of forecast and real data for several days (from 15.04.20 to 21.04.20) for the same four scenarios for the development of the epidemic in Moscow made on April 7. Tables 7 and 8 show that the milder scenarios of the epidemic in Moscow (the third and fourth), unfortunately, could not be kept. The best description of statistical data on coronavirus cases in Moscow in the last days are given by the 1-st and 2-nd scenarios with a normalizing factors $N = 1.0 \times 10^6$ and $N = 7.6 \times 10^5$. In the last week, Moscow has sharply tightened quarantine measures and increased control over their implementation. This suggests optimism that the second scenario of the epidemic will be held. But there is a chance that Moscow will follow the worst-case first scenario.

**Table 7. Calculated data of peculiar parameters of the COVID-19 coronavirus epidemic in Moscow under 4 scenarios**

| № sc. | Normalizing factor $N$ | Peak date | Cases at the peak | Daily cases at the peak | Total number of cases | End time of the epidemic |
|---|---|---|---|---|---|---|
| 1 | $1 \times 10^6$ | 24.04 | 38152 | 2959 | 74073 | mid-late June |
| 2 | $7.6 \times 10^5$ | 22.04 | 29929 | 2359 | 57590 | early June |
| 3 | $5 \times 10^5$ | 19.04 | 20831 | 1646 | 39168 | early June |
| 4 | $3 \times 10^5$ | 15.04 | 13848 | 1235 | 26027 | end of May |

**Table 8. Comparison of real and calculated data for Moscow**

| Day | Reported cases | 1 scenario: $N = 1 \times 10^6$ | 2 scenario: $N = 7.6 \times 10^5$ | 3 scenario: $N = 5 \times 10^5$ | 4 scenario: $N = 3 \times 10^5$ |
|---|---|---|---|---|---|
| April 14 | 13002 | 13310 | 13140 | 12848 | 12612 |
| April 15 | 14776 | 15101 | 14839 | 14336 | 13848 |
| April 16 | 16146 | 17070 | 16680 | 15898 | 15074 |
| April 17 | 18105 | 19215 | 18656 | 17515 | 16269 |
| April 18 | 20754 | 21535 | 20753 | 19167 | 17413 |
| April 19 | 24324 | 24018 | 22954 | 20832 | 18490 |
| April 20 | 26350 | 26650 | 25325 | 22483 | 19485 |
| April 21 | 29433 | 29408 | 27570 | 24100 | 20393 |

Now let's look at the dynamics of the spread of COVID-19 coronavirus infection in regional Russia (without Moscow and the Moscow region). When calculating the number of people infected with COVID-19 coronavirus for Russian regions, we also considered four scenarios with the ratio of normalized multipliers to the number of Russia (minus the number of the Moscow region) in the range [0.8-1.7%], namely, scenarios with normalized multipliers:

1$^{st}$ scenario: $N = 2 \cdot 10^6$; 2$^{nd}$ scenario: $N = 1 \cdot 10^6$;

3$^{rd}$ scenario: $N = 7.6 \cdot 10^5$; 4$^{th}$ scenario: $N = 5 \cdot 10^5$.

Table 5 shows that scenarios 3 and 4 produce results that lag more and more behind the actual data every day. So, it was not possible to keep the milder scenarios of the epidemic development with a smaller number of cases. In recent days, the results of scenarios 1 and 2 have also started to lag behind the real data. It seems that the easing of already soft restrictive measures has led to an outbreak in some regions. This means that the growth rate parameter $\alpha$ has changed, and it will need to be slightly increased and selected in the model. Which of the scenarios, the harder 1$^{st}$ scenario or the less hard 2$^{nd}$ scenario, is implemented will be clear in a few days.

If you compare the growth in the number of infected people in Russia with the growth in the number of infected people in Moscow, we can see that the indicators for the Russian regions is generally lower than for Moscow. The average population growth rate for Moscow, calculated before the peak, is 1.112, which is lower than in European countries, but higher than in Japan (see Table 1). The average number of COVID-19 infected people (before the peak of the epidemic) in Russian regions is 1.098, which is lower than in Moscow. That is, the growth rate of the number of COVID-19 infected people in the year Moscow is 1.02 times higher than the growth rate of the number of cases in the Russian regions (in total).

The epidemic peaks in Moscow under various scenarios occur in the range from 19.04 to 24.04. and the epidemic peaks in Russian regions for various scenarios from 25.04 to 4.05. In other words, the epidemic peaks in the Russian regions are a week behind the epidemic peaks in Moscow.

The forecast was made with the indicator value unchanged from April 1. And the value of this parameter is correlated with the number of people in contact. We see that the measures taken to isolate the population have had an effect, and in all scenarios the growth rate has decreased from April 1.

It should be noted that the distribution of the epidemic in China fell well within the scenario with the choice of a single normalizing factor. To model the epidemic in a number of European countries, such as Italy and Spain, we needed two logistic equations with different rationing factors. This is due to fluctuations with a large amplitude in the peak area and on a downtrend. The results of the logistics equation with a smaller multiplier described the actual data better before the peak. The model with a larger multiplier described the real data better after the peak. In other words, the actual data went along the corridor between these two curves.

What is the scenario for the spread of the epidemic in Russia after the "Chinese" or "European" peak will become clear when the peak will be passed.

**Conclusions**

Mathematical modeling of the dynamics of the spread of COVID-19 coronavirus infection in Moscow and in regional Russia was carried out and forecasts were made. The simplest nonlinear discrete equation describing logistic growth was chosen as the model. This model contains only two parameters that are chosen based on statistical data at the initial stage of the epidemic. The first parameter is a rate of the increase in the number of cases. It can change with the implementation of quarantine measures and affect on the total number of cases at the end of the epidemic. The second parameter is a normalization multiplier that only estimates the maximum potential number of residents who may become ill. The predicted total number of cases is calculated in the model and may differ from the N value by more than an order of magnitude. Its exact value can only be determined closer to the peak of the epidemic, so based on the experience of modeling the spread of the epidemic in different countries that have already passed the peak, different N values are selected and different scenarios are considered. Four different scenarios of the epidemic development are proposed for Moscow and regional Russia. For each scenario, the growth rates of the population infected with COVID-19 coronavirus were founded, the peak time was calculated, the maximum increase in cases at the peak and the total number of cases during the epidemic were found. While the work was being prepared, it became clear that Moscow did not manage to isolate citizens well from each other, and two easier scenarios are no longer being implemented. The tightening of quarantine measures in the last week inspires optimism that the development of the epidemic in Moscow will follow the second, not the worst scenario, which currently best describes the actual data.

Mathematical modeling of the dynamics of the spread of coronavirus infection in the regions of Russia showed that it is delayed by 6 days compared to Moscow, and the growth rates of the number of cases is slightly less than the growth rate in Moscow. Four scenarios for the development of the epidemic in regional Russia were considered too. It is calculated that for the" light "scenario for the Russian regions, the number of cases will be ~ 40,000 people, for the "heavy" scenario – 160,000 people.

Note that we based on official data on the number of cases, which determine the burden on health care. And unrecorded cases, which according to some estimates may be 20% of those accounted for, ensure the spread of infection, and the model can be used.

**Acknowledgements:** The authors are grateful to the postgraduate student A.S. Shaneva of D. Mendeleev University for help in editing the article.